\documentclass[final,1p,times]{elsarticle}

\usepackage{graphicx}
\usepackage{amssymb}
\usepackage{amsthm}
\usepackage{lineno}

\usepackage{graphics}
\usepackage{color}

\journal{Nuclear Physics A}

\begin{document}

\begin{frontmatter}

\title{QGP viscosity at RHIC and the LHC - a 2012 status report}

\author{Huichao Song}

\address{Department of Physics, Peking University, Beijing 10081, China \\
Department of Physics, The Ohio State University, Columbus, Ohio 43210-1117, USA}



\begin{abstract}

In this article, we briefly review recent progress related to extracting the quark-gluon plasma (QGP) specific shear viscosity from the flow data measured at Relativistic Heavy-Ion Collider (RHIC) and the Large Hadron Collider (LHC).

\end{abstract}

\end{frontmatter} 


\section{Introduction}
Heavy ion experiments at the Relativistic Heavy- Ion Collider
(RHIC) at BNL and the
Large Hadron Collider (LHC) at CERN have accumulated
strong evidences for the creation of the quark gluon
plasma (QGP) that is strongly coupled and behaves like an almost perfect liquid. The transport properties of the QGP fluid is a currently hot topic. Due to the difficulties of the first principle calculations, it is desirable to extract them from experimental data. In this proceeding, we will review recent progress related to extracting the QGP shear viscosity from the flow data at RHIC and the LHC.

\section{Hydrodynamics and hybrid models}

Viscous hydrodynamics is an useful tool to study the viscous effects on the QGP fireball evolution and final observables for relativistic heavy-ion collisions at RHIC and the LHC energies. During the past years, several groups have independently developed (2+1)-d~\cite{Song:2007fn,Romatschke:2007mq,Dusling:2007gi} and (3+1)-d~\cite{Schenke:2010rr,Bozek:2011ua} viscous hydrodynamic codes with/without longitudinal boost invariance. A recent comparison between 2+1-d and 3+1-d viscous hydrodynamics shows that the realistic longitudinal expansion from the 3+1-d code only slightly affects the flow at mid-rapidity~\cite{ChunNote}. Therefore, one can safely implement the 2+1-d code to investigate the soft physics at mid-rapidity.

With the efforts from different viscous hydrodynamic groups,   flow has been widely accepted as the key observable to extract the QGP viscosity. However, the hadronic chemical composition and non-equilibrium kinetics also play important roles for flow development in the hadronic phase, bringing additional uncertainties for the extracted QGP viscosity.  For a better description of the hadronic matter, one further developed the viscous hydrodynamics + hadron cascade hybrid approach that connects the hydrodynamic evolution for a viscous QGP fluid to the microscopic description for the re-scatterting and decoupling of the late hadronic matter. Such hybrid approaches include {\tt VISHNU} hybrid model~\cite{Song:2010aq}  developed in 2010 based on 2+1-d viscous hydrodynamics and UrQMD hadron cascade (see Sec 4 for a summary of the main results), the McGill code developed in this year based on 3+1-d viscous hydrodynamics and UrQMD (the new results can be found in Ref.~\cite{Jeon}) and the Livermore hybrid code (see Sec.7 and references therein). \\[-0.10 in]

\underline{Single shot vs. event-by-event simulations:} Many hydrodynamic calculations concentrate on single-shot simulations for computing efficiency. Recently, several groups extended single-shot hydrodynamic simulations to event-by-event ones, making it possible to investigate the hydrodynamic response of the initial state fluctuations ~\cite{Schenke:2010rr,Qiu:2011iv,Qin:2010pf,Qiu:2011fi}. Single shot simulations with smooth initial conditions obtained by averaging over large numbers of fluctuating events after rotation to align the ``event planes" are, in principle, a computationally efficient short-cut to investigate initial-state fluctuation effects on even and odd
flow coefficients. However, a detailed comparison between event-by-event and single-shot hydrodynamics shows O(10\%) deviations for the elliptic and triangular
flows, especially when the QGP viscosity is small~\cite{Qiu:2011fi}.
Furthermore, the higher order flow harmonics $v_4, v_5$ and $v_6$ can not be reliably computed within a single-shot hydrodynamic approach due to the mode-coupling effects. With the appearance of new flow measurements (event-by-event $v_n$ distribution, event plan correlation between flow angles, etc., please refer to Sec.7 for details) and the ever increasing computing and storage capabilities, the transition to event-by-event simulations are inevitable.

\section{The QGP shear viscosity from the elliptic flow data -- an early attempt}
The elliptic flow and higher order flow harmonics are important observables for the bulk matter created in relativistic heavy ion collisions. It was found that the elliptic flow $v_2$ and triangular flow $v_3$ are very sensitive to the shear viscosity. Even the conjectured lowest value $\eta/s=1/(4\pi)$ from the AdS/CFT correspondence leads to a significant suppression of  $v_2$ and $v_3$~\cite{Song:2007fn,Romatschke:2007mq,Dusling:2007gi,Schenke:2010rr}.  The bulk viscosity also suppress the development of flow anisotropy. Its effects are much smaller than the shear viscous ones due to the critical slowing down near the phase transition.  Therefore, one can extract the QGP shear viscosity from flow data without large contaminations from the bulk viscosity~\cite{Song:2009rh}.

The first attempt to extract the QGP viscosity from the elliptic flow data, using 2+1-d viscous hydrodynamics, was done by Luzum and Romatschke around 2008~\cite{Romatschke:2007mq}. They implemented two initial conditions from optical Glauber and KLN models and found that the $\sim$$30\%$ uncertainties in the initial eccentricity lead to $\sim$$30\%$ uncertainties for the calulated elliptic flow, which translates into $\sim$$100\%$ uncertainties for the extracted value of the QGP shear viscosity. The effects neglected in this work are the off-equilibrium kinetics (or so-called highly viscous hadronic effects) and the partially chemically equilibrated nature of the hadronic phase,  and the initial state fluctuations.  Around that time, the effects from bulk viscosity was unknown.   After making generous estimations for all of these uncertainties, it appears that the averaged specific QGP shear viscosity, cannot exceed
the following conservative upper limit~\cite{Romatschke:2007mq,Song:2008hj}:
\begin{eqnarray*}
 \left.\frac{\eta}{s}\right|_\mathrm{QGP} < 5\times \frac{1}{4 \pi}.
\end{eqnarray*}

\section{The QGP shear viscosity at RHIC and LHC energies -- the current status}

\subsection{The QGP viscosity at RHIC energies}
With {\tt VISHNU} hybrid model becoming available and the effects from hadronic matter being in control,  we were ready to extract the QGP shear viscosity from the flow data with reliable uncertainty estimates.  Ref.~\cite{Song:2010mg} proposes to extract the QGP viscosity from the integrated $v_2$ data for all charged hadrons since it is most directly related to the fluid momentum anisotropy and less sensitive to the details of {\tt VISHNU} calculations, such as the form of the non-equilibrium distribution function $\delta f$, the bulk viscosity, initial flow, etc~\cite{Song:2010mg,Song:2011hk}. It was found that the eccentricity-scaled elliptic flow $v_2^{ch}/\varepsilon_x$ as a function of charged multiplicity per unit overlap area $\mathrm{(1/S)dN_{ch}/dy}$ is universal that depends only on the QGP shear viscosity but not on the
initialization models~\cite{Song:2010mg}. It thus desirable to use these curves to extract the QGP viscosity.  Fig.~1 shows a comparison of the theoretical and experimental $v_2^{ch}/\varepsilon_x-\mathrm{(1/S)dN_{ch}/dy}$ curves, where  the left and right panels correspond to MC-Glauber and MC-KLN  initialization models. The theoretical lines are calculated from single shot {\tt VISHNU} simulations with different $(\eta/s)_{QGP}$ as inputs, using event-averaged smooth initial conditions generated from these two models. The experimental data  are the corrected elliptic flow with non-flow and fluctuation effects removed~\cite{Ollitrault:2009ie}.
Due to the $\sim$$20\%$ larger ellipticity of the MC-KLN
fireballs, the magnitude of $v_2^{exp}/\varepsilon_x$ differs between the two models. As a result, the extracted value of $(\eta/s)_{QGP}$ from these two panels changes by a factor of 2. Taken the main uncertainties from these two initial conditions, we found that {the averaged specific
shear viscosity of the QGP created at top RHIC energies is~\cite{Song:2010mg}:}
\begin{eqnarray*}
\frac{1}{4 \pi} <\left.\frac{\eta}{s}\right|_\mathrm{QGP} < 2.5\times \frac{1}{4 \pi}
\end{eqnarray*}

\underline{Remarks:} Small bulk viscous effects and proper event-by-event hydrodynamical evolution of fluctuating initial conditions slightly reduce the integrated $v_2$. while early flow slightly increase it. Although they should be studied in more quantitative detail, we expect the total uncertainty band translated to the above extracted value of QGP shear viscosity may slightly shift after cancelations. For more details on these uncertainty estimates, please refer to~\cite{Song:2012tv}. \\[-0.10in]

\begin{figure*}[t]
\vspace*{0.0cm}
\center
\resizebox{0.70\textwidth}{4.5cm}{%
  \includegraphics{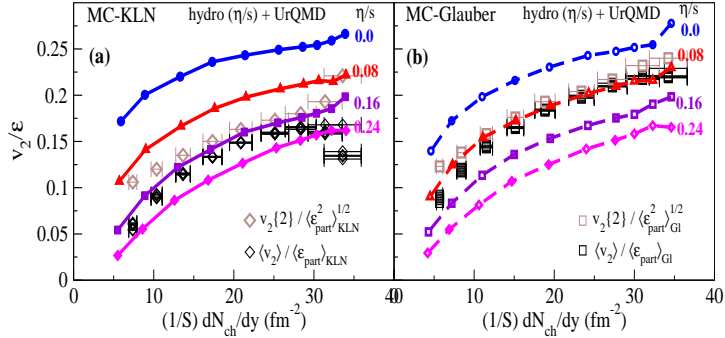}}
\vspace*{0cm}  
\caption{(Color online) eccentricity-scaled elliptic flow as a function of final multiplicity per
 area.}
\end{figure*}

With $(\eta/s)_{QGP} \simeq (1/4\pi)$ for MC-Glauber and $(\eta/s)_{QGP} \simeq (2/4\pi)$ for MC-KLN extracted from the
$p_T$ integrated $v_2$ for all charged hadrons, {\tt VISHNU} yields an excellent description of the $p_T$-spectra and differential elliptic flow $v_2(p_T)$ (for $p_T <2 GeV$)  for all charged hadrons and identified hadrons at different centrality bins measured in 200 A GeV Au+Au collisions at RHIC~\cite{Song:2011hk}, showing the power of the {\tt VISHNU} hybrid model and the robustness of the above extraction of the QGP shear viscosity.

\subsection{Recent results at the LHC}

\underline{Elliptic flow and the QGP shear viscosity at the LHC:} After extrapolating our calculation to the 2.76 A TeV Pb+Pb collisions at the LHC, we found that the $(\eta/s)_{QGP}$ extracted from RHIC slightly over-predicts the LHC elliptic flow data. After increasing  $(\eta/s)_{QGP}$ from $\sim 2/(4\pi)$ to $\sim 2.5/(4\pi)$ (for MC-KLN)\footnote{For computing efficiency, we only investigate the case with MC-KLN initializations in details. The trend on how $(\eta/s)_{QGP}$ changes from RHIC to LHC energies is similar for MC-Glauber and MC-KLN initializations~\cite{Song}.}, {\tt VISHNU} gives a better description of the integrated and differential elliptic flow data for all charged hadrons at the LHC~\cite{Song:2011qa}.
However, this dost not necessarily mean that the QGP turns more viscous at higher temperatures reached by LHC since $v_2$ becomes sensitive to the initial shear stress tensor at the LHC and one needs to further investigate the temperature-dependent QGP viscosity (please refer to Sec.~6 for detailed discussions).

The ALICE collaborations also measured the differential elliptic flow $v_2(p_T)$ for identified hadrons from central to peripheral collisions. Although purely viscous hydrodynamic simulations makes a nice description of the pion and kaon data at various centralities, it over-predicts the proton $v_2(p_T)$ from most central to semi-central collisions~\cite{Shen:2011eg,Heinz:2011kt}. This problem was fixed by the later {\tt VISHNU} hybrid model simulations. With the microscopic rescattering of the hadronic matter, the generations of radial and elliptic flow for protons are re-balanced. Besides keeping the nice descriptions of the pion and kaon data, {\tt VISHNU} gives an improved description of the proton elliptic flow data at various centrality bins, especially for the most central to semi-central collisions~\cite{Heinz:2011kt}\\[-0.10in]

\underline{A simultaneous fitting of the elliptic and triangular flow data:}  The QGP shear viscosity suppresses the development of both elliptic flow $v_2$ and triangular flow $v_3$~\cite{Song:2007fn,Romatschke:2007mq,Dusling:2007gi,Schenke:2010rr,Qiu:2011fi}. While the initial eccentricities $\varepsilon_2$ differ by O(20\%) between MC-KLN and MC-Glauber models, the triangular deformation $\varepsilon_3$ are almost identical.  As a result, $v_3$ is much less sensitive to these two initializations when compared with $v_2$~\cite{Qiu:2011fi}. Using pure viscous hydrodynamics, a simultaneous fitting of the elliptic and triangular flow data indicates that these recent LHC data favors a small value of QGP shear viscosity around $1/(4\pi)$. With that value, the centrality dependent integrated  elliptic and triangular flow
are well described by the MC-Glauber initial conditions, while MC-KLN initial conditions obviously under-predict the triangular flow data with the QGP viscosity tuned to fit the elliptic flow~\cite{Qiu:2011hf}.  Although more comprehensive investigations from  the computing expensive {\tt VISHNU} hybrid model are needed in the near future, the hadronic afterburner will not change the main conclusion since both elliptic and triangular flow almost reach saturation in the QGP phase at the LHC energies~\cite{hirano-QM2006}. \\[-0.10in]

\underline{The proton puzzle at the LHC:} Recently, ALICE collaborations measured the $\bar{p}/\pi^-$ ratio in 2.76 A TeV Pb+Pb collisions at the LHC~\cite{ALICE}, which obviously disagrees with the value observed by STAR in 200 A GeV Au+Au collisions at RHIC, but agrees with the one from PHENIX. It was found that $\bar{p}/\pi^-$ ratio is sensitive to the chemical freeze-out temperature $T_{chem.}$ in theoretical calulations. Correspondingly, hydrodynamic simulations with $T_{chem.}=165 \ \mathrm{MeV}$ (chosen the value measured at RHIC) over-predict the $\bar{p}$ spectra at the LHC~\cite{Shen:2011eg}. Our past {\tt VISHNU} simulations (that accidentally tuned off the baryon anti-baryon  $B-\bar{B}$ annihilation channels) with the switching temperature $T_{sw}=T_{chem.}=165 \ \mathrm{MeV}$  also over-predict the $\bar{p}$ spectra at the LHC~\cite{Song:2011qa}. To solve this problem, some groups proposed to use a lower chemical freeze-out temperature to fit the data. However, new {\tt VISHNU} simulations with the $B-\bar{B}$ annihilation channels shows that such annihilations reduce the proton and anti-proton multiplicity by about 30\% in central collisions and by about 15\% in peripheral collisions, leading a dramatically improved description of the proton and anti-proton spectra measured at the LHC~\cite{song2012}. With that $B-\bar{B}$ annihilation, the descriptions of the proton and anti-proton final-multiplicity and spectra at RHIC (measured by PHENIX) are also improved~\cite{song2012}. Although the detailed work is still on-going, this helps to explain the theoretical over-prediction of the proton, anti-proton spectra, and $\bar{p}/\pi^-$ ratio at the LHC with  the chemical freeze-out temperature set to 165 MeV.

\section{Recent developments on initialization models}
The MC-Glauber and MC-KLN models, commonly used by many groups, construct the initial state fluctuations through the positions fluctuations of the nucleons inside the colliding nuclei. Recently, several groups further investigate the quantum fluctuations of color charges. Combing the classical Yang-Mill's approach for the Glasma field with the impact parameter dependent saturation model (called IP-Glama model), Schenke, Tribedy and Venugopalan studied the color charge fluctuations and found that such effects moderately change the initial eccentricities compared with the one from traditional MC-KLN model. Ref.~\cite{Muller:2011bb} investigated the transverse correlations for energy density fluctuations within the framework of Color Glass Condensate (CGC).
Based on this work, the OSU group constructed a new Monte-Carlo initialization generator to produce initial energy density profiles with correlated fluctuations and found a small increase of $\varepsilon_2-\varepsilon_5$ \cite{Ulrich}. In Ref.~\cite{Dumitru:2012yr}, the fluctuations of the initial gluon production\footnote{For other models including multiplicity fluctuations for soft particle productions, please refer to Ref.~\cite{Qin:2010pf,URQMD,AMPT,HIJING}.} are investigated accounting to a negative binomial distribution within the $k_T$ factorization approach of CGC. It gives an initial eccentricity  $\varepsilon_2$ close to the one from traditional MC-KLN model, but obvious larger $\varepsilon_3 -\varepsilon_5$ than the MC-KLN one. The larger higher order eccentricity leads to larger value of theoretical higher order flow harmonics, which requires a significantly larger value of the QGP shear viscosity to fit the corresponding experimental data. This brings additional uncertainties for our extracted $(\eta/s)_{QGP}$  from the elliptic flow data alone. A joint investigation of elliptic, triangular and higher order flow harmonics can help to further constrain the initialization models and yield a more accurate value for $(\eta/s)_{QGP}$.

Hydrodynamic initial conditions can also be provided by dynamical models, such as URQMD \cite{URQMD}, EPOS~\cite{EPOS}, AMPT~\cite{AMPT}, IP-Glama~\cite{Schenke:2012wb}, etc., which try to account for the pre-equilibrium dynamics before thermalization. Generally, pre-equilibrium dynamics reduces the initial eccentricities $\varepsilon_2$ and $\varepsilon_3$ and contributes radial and anisotropic flow. It was found that the additional fluctuations from the energy deposition in UrQMD slightly increase $\varepsilon_2$ and $\varepsilon_3$~\cite{URQMD} and the early longitudinal flow fluctuations from AMPT slightly reduce $v_2$~\cite{AMPT}. Except for IP-Glasma, most of these dynamical models are matched to ideal hydrodynamics for further evolution. How they quantitatively influence the extracted value of the QGP shear viscosity is still unknown and needs additional work.

\section{Temperature-dependent $(\eta/s)_{QGP}(T)$}

Besides further increasing the precision of the extracted ``averaged" specific QGP shear viscosity, it is important to quantitatively determine the temperature-dependent  $(\eta/s)_{QGP}(T)$ to further explore the fundamental properties of the QCD matter at high temperatures.
Past research showed that one needs a slightly larger constant  $\eta/s$ to fit the LHC flow data than the one used for RHIC~\cite{Song:2011qa,Schenke}, which demonstrates  that it is feasible to extract a temperature-dependent $(\eta/s)_{QGP}(T)$ from the new measured flow data at different collision energies. One of the crucial issues is to extract the initial temperature since it can not be directly measured in experiments. The new developed massive data evaluating technique~\cite{Soltz:2012rk} make it possible to simultaneous extract the QGP shear viscosity, the initial temperature and other parameters in initial conditions from multiple data sets. The error bands of these extracted values are largely controlled by the precision of the measured experimental data.  It is thus desirable to further reduce the systematic and statistical uncertainties of related observables (such as final multiplicities and the $p_T$ spectra, HBT radii, flow harmonics $v_n$, event plan correlations of flow angles, $v_n$ distributions, etc., please also refer to Sec.~7 for details) for a quantitatively determination of $(\eta/s)_{QGP}(T)$.

\section{Other related current developments}

\underline{Event plane correlations for flow angles:} Recently ATLAS collaborations measured the event plan correlations between the flow angles associated with higher order flow harmonics in 2.76 A TeV Pb+Pb collisions~\cite{ATLAS}. Event-by-event hydrodynamic simulations show that the correlation strength is sensitive to both the initial conditions and the QGP shear viscosity~\cite{Qiu:2012uy}. Although more detailed studies are needed for a quantitative description of the data, it is impressive that e-by-e hydrodynamics correctly reproduced the qualitative features of the centrality dependent correlations between different flow angles, providing strong support for the fluid dynamic description of the fireball evolution.\\[-0.1 in]

\underline{Event-by-event distribution of $v_n$:}  ATLAS also measured the event-by-event distributions of $v_2-v_4$ at varies centralities in 2.76 A TeV Pb+Pb collisions~\cite{Jia}. The event-by-event hydrodynamic simulations with MC-Glauber and MC-KLN initializations show that none of these two initializations works for the distributions for all $v_n$~\cite{Heinz}. In contrast, IP-Glasma gives the $\varepsilon_n$ distributions that largely overlap with these measured $v_n$ distributions. After the hydrodynamic evolution, the descriptions of the data are further improved~\cite{Schenke}. A future study of these flow distributions with different initializations as inputs may help us to constrain the initialization models and to understand the sources of fluctuations.  \\[-0.1 in]

\underline{Higher order flow harmonics in ultra-central collisions:} In Ref.~\cite{Luzum}, Luzum proposed to extract the QGP viscosity from a simultaneous fitting of all measured $v_n$ in the ultra-central collisions since the fluctuation effects are dominated and the geometry effects are suppressed there. However, viscous hydrodynamic simulations from the OSU group, with different $\eta/s$ as input, show that neither the MC-KLN nor the MC-Glauber model appears to be able to simultaneously describe all measured $v_n$ from CMS~\cite{Qiu}. It will be interesting to see a comparison of the otherwise very successful IP-Glasma model~\cite{Schenke} with these ultra-central Pb-Pb data in the near future. \\[-0.1 in]

\underline{Systematic $\chi^2$ fitting of the experimental data:} Using a hybrid model that connect {\tt UVH2+1D} viscous hydrodynamics with {\tt UrQMD} hadron cascade model, Soltz and his collaborators made a systematic $\chi^2$ evaluation of the  pion spectra, elliptic flow and HBT radii measured by STAR and PHENIX in 200 A GeV Au+Au collisions~\cite{Soltz:2012rk}.  The evaluation is performed in a two-dimensional parameter space for initial temperature and the QGP shear viscosity. For each inputting initial conditions (with/without pre-equilibrium flow),
the fitted value for them is give by the lowest value of the sum of $\chi^2$ for these three observables.  Although this work only concentrate on the data sets in the 0-20\% centrality bin and has lots of room for future improvement, it demonstrates a systematic evaluation technique for constraining the multiple theoretical parameters from the multiple experimental data sets for relativistic heavy ion collisions.

\begin{figure*}[t]
\vspace*{0.0cm}
\center
\resizebox{0.80\textwidth}{4.7cm}{%
  \includegraphics{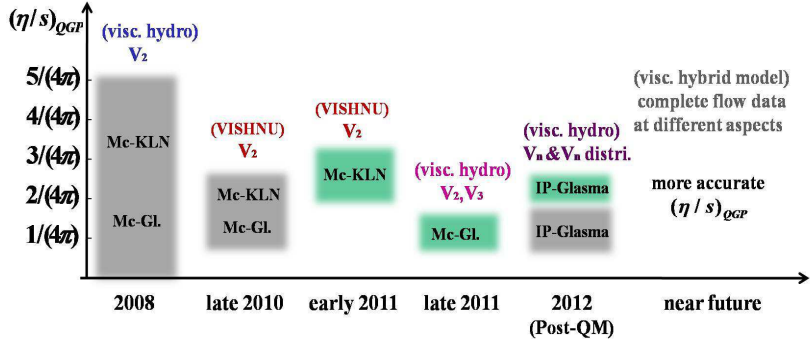}}
\vspace*{0cm}  
\caption{(Color online) Time line for the extracted QGP specific shear viscosity with uncertainty estimations. Grey bands are for RHIC (200 A GeV Au+Au collisions) and dark green band are for LHC (2.76 A TeV Pb+Pb collisions). }
\end{figure*}

\section{Summary and concluding remarks}
In summary, our field has evolved quickly during the past few years. Using pure viscous hydrodynamics, an early extraction of the QGP shear viscosity from elliptic flow data suggested $(\eta/s)_{QGP}<5/(4\pi)$~\cite{Romatschke:2007mq}, with the main uncertainties coming from the undetermined initial conditions, an improper description of hadronic chemical composition and non-equilibrium kinetics of the late hadronic stage.  With the hadronic afterburner that realistically describes these two hadronic effects, the {\tt VISHNU} hybrid model fits the centrality-dependent integrated elliptic flow data with MC-Glauber and MC-KLN initial conditions and gives $1/(4\pi)<(\eta/s)_{QGP}<2.5/(4\pi)$, where the uncertainty is now entirely dominated by the different initial eccentricities from these two models~\cite{Song:2010mg}. One also found that the averaged QGP shear viscosity at the LHC is close to (or slightly larger than) the one at RHIC~\cite{Song:2011qa}. Using pure viscous hydrodynamics, a simultaneous fit of the elliptic and triangular flow at 2.76 A TeV Pb+Pb collisions strongly indicated that these data prefer an even smaller value of the specific QGP viscosity around  $1/(4\pi)$. With that value, the MC-Glauber initial conditions can fit the centrality dependent elliptic and triangular flow data, whereas MC-KLN initial conditions fail to do so~\cite{Qiu:2011hf}.  Since both elliptic and triangular flow almost reach saturation in the QGP phase at the LHC, future sophisticated {\tt VISHNU} calculations with the hadron afterburner are not expected to change these conclusions. The MC-Glauber and MC-KLN initial condition models used by many hydrodynamic simulations (including our own) construct the initial state fluctuations from nucleon position fluctuations inside the colliding nuclei. Recently, several groups further studied other sources of quantum fluctuations in the initial state, such as color charge fluctuations, multiplicity fluctuations, initial flow fluctuations, etc.~\cite{Schenke:2012wb,Muller:2011bb,Ulrich,Dumitru:2012yr,AMPT}. Initially this led to large additional uncertainties for the extracted QGP shear viscosity. A recent fit of $v_n$ and event-by-event $v_n$ distributions with 3+1-d viscous hydrodynamics using IP-Glasma initial conditions \cite{Schenke} dramatically reduced these uncertainties, by eliminating both the MC-Glauber and MC-KLN models as viable models for the initial-state fluctuations \cite{Schenke,Heinz}. The combined analysis of all measured $v_n$ ($n=1,\dots,5$) gives $\eta/s \sim 0.20$ for LHC collisions and the smaller value $\eta/s \sim 0.12$ for RHIC collisions \cite{Schenke}.  Although detailed studies of initialization models are still on-going, additional systematic anisotropic flow studies (including the centrality dependence of all integrated and differential flow harmonics for all charged and identified hadrons, event plane correlations between flow angles, flow distributions, etc.) are expected to strongly constrain the initial conditions and to further narrow the error band for the extracted QGP viscosity. Progress is happening quickly, and exciting new results are expected to be available by the next Quark Matter meeting.

\newpage

\emph{Acknowledgement:} The author thanks for discussions with S.~A. Bass, U.~Heinz, P.~Huovinen, J.~Jia, M.~Luzum, Z.~Qiu, L.~G.~Pang, H.~Peterson, S.~Moreland, B.~Schenke, C.~Shen, and X.~N.~Wang. This work was supported by the U.S. Department of Energy under Grants No. \rm{DE-SC0004286} and (within the framework of the JET Collaboration) \rm{DE-SC0004104}.


\end{document}